\documentclass[preprint2]{aastex}%

\usepackage{graphicx}
\usepackage{amsmath}
\usepackage{amssymb}
\usepackage{epsfig}
\usepackage{color}
\usepackage{bm}

\graphicspath{{Pics/}}
\shorttitle{Shock formation in e-i plamas}
\shortauthors{Stockem et al.}

\begin{document}

\title{Shock formation in electron-ion plasmas: mechanism and timing}

\author{A. Stockem Novo}
\affil{Institut f\"ur Theoretische Physik, Lehrstuhl IV: Weltraum- \& Astrophysik, Ruhr-Universit\"at, Bochum, Germany}
\email{anne@tp4.rub.de}

\author{A. Bret}
\affil{ETSI Industriales, Universidad de Castilla-La Mancha, 13071 Ciudad Real, Spain\\
Instituto de Investigaciones Energ\'eticas y Aplicaciones Industriales, Campus Universitario de Ciudad Real, 13071 Ciudad Real, Spain}

\author{R. A. Fonseca\altaffilmark{1} and L. O. Silva}
\affil{GoLP/Instituto de Plasmas e Fus\~ao Nuclear, Instituto Superior T\'ecnico, Universidade de Lisboa, Lisboa, Portugal}
\altaffiltext{1}{also at DCTI, ISCTE - Lisbon University Institute Portugal}

\begin{abstract}
We analyse the full shock formation process in electron-ion plasmas in theory and simulations. It is accepted that electromagnetic shocks in initially unmagnetised relativistic plasmas are triggered by the filamentation instability. However, the transition from the first unstable phase to the quasi-steady shock is still missing. We derive a theoretical model for the shock formation time, taking into account the filament merging in the non-linear phase of the filamentation instability. This process is much slower than in electron-positron pair shocks, so that the shock formation is longer by a factor proportional to \(\sqrt{m_i/m_e} \ln(m_i/m_e)\).
\end{abstract}

\keywords{instabilities --- relativistic processes --- shock waves}


\section{Introduction}

Collisionless shocks are ubiquitous in astrophysical environments, such as gamma-ray bursts, active galactic nuclei or pulsar wind nebulae. They are especially important in the context of cosmic ray acceleration. Once the shock is in a quasi-steady state, the jump conditions can be determined from the conservation of mass, energy and momentum in a fluid model \citep[][]{BM76}. The density jump from the upstream to the downstream of a relativistic strong shock is given in 2D by \(n_2/n_1 \approx 3\). Plasma instabilities are the mediators of such collisionless shocks \citep[][]{S66}, but it is still not known precisely how the transition from the initial plasma turbulence to the final quasi-steady state of the shock happens.

In a symmetric counter-streaming flow of charged particles, plasma instabilities develop, causing an isotropisation of the particle momenta and initiating the deceleration of the plasma flows. In relativistic, cold and initially unmagnetised plasmas, electromagnetic current filamentation modes are dominant which inhabit a strong perpendicular component \citep[][]{W59,F59}. The energy is transferred from an initial longitudinal streaming into the perpendicular directions. The particle forward motion is slowed-down and a collisionless shock starts to form with a density ratio of \(n_2/n_1>2\).

Particle-in-cell (PIC) simulations are an excellent tool to investigate the shock formation process, since non-linear processes are involved \citep[][]{MedvedevApJ2005,HH08,Spitkovsky2008,MF09,NN11,FF12,SS13,SF14,HF15}. In a previous study, we identified the shock formation time of electron-positron pair shocks as twice the saturation time of the magnetic field amplification due to the filamentation instability, \(\tau_{f,e} = 2 \tau_{s,e}\) \citep[]{BS13,BS14}. Surprisingly, electron-ion shocks do not show the same feature as one would expect from the rapid relativistic mass increase of the electrons. Our analysis shows a shock formation time \(\tau_{f,i} \approx 3 \, \sqrt{m_i/m_e}  \, \tau_{f,e}\). However, a similarity between both scenarios was observed: the steady state shock is formed when the first ions start to recirculate. The non-linear phase is faster in pair shocks and no additional time is necessary for the merging of magnetic flux tubes, so that the recirculation process starts earlier. As a consequence, particles are mainly scattered in the turbulence rather than doing a full gyro rotation, which is why the isotropisation process is less effective. We present a detailed theory for the shock formation in electron-ion plasmas, and compare it to state-of-the-art PIC simulations.

\section{Shock formation time}\label{shockform}
We consider a simple scenario for shock formation with the plasma initially being unmagnetised and symmetric counterstreaming relativistic beams of electrons and ions of mass ratio \(m_i\geq m_e\) and Lorentz factor \(\gamma_0\). We refer to pair shocks of electrons and positrons when \(m_i = m_e\). The beams are initially cold, characterised by the temperature parameter \(\mu = \gamma_0 mc^2 / k_B T\), where the mass \(m\) and temperature \(T\) refer to the respective species. Such scenarios are a stimulating environment for the current filamentation instability \citep[][]{F59} to occur. The growth rate of the cold current filamentation instability, for electron beams as well as for ion beams, is given by \(\delta = \sqrt{2/\gamma_0}\beta_0 \omega_{p}\), where \(\beta_0 = v_0 / c\) denotes the normalised fluid velocity of the beam and the plasma frequency is given by \(\omega_{p} = \sqrt{4\pi n_0 e^2/m}\) with the initial uniform beam density \(n_0\) \citep{BretPoPReview}.

Let us start describing the mechanisms at work that lead to shock formation, before we turn to the analytical evaluation of the shock formation time. As already highlighted by various authors \citep{lyubarsky06,Shaisultanov2012,Davis2013}, the electron Weibel instability is first triggered when the two plasmas start overlapping. By the time it saturates, it has generated filaments of the size of the electron Larmor radius in the field at saturation \citep{BS13}. At the time the ion Weibel instability starts to grow, an unstable wavelength has already been seeded. It corresponds precisely to the one that has resulted from the merging of the filaments associated with the electron Weibel instability with a typical length scale of \(c/\omega_{pe}\). As a consequence, by the time the ion Weibel instability saturates, the field is near equipartition with the ions, but the filaments are still the size of the \emph{electronic} Larmor radius, not the ions one, since the instability on the ions was initially seeded at these length scales. Such filaments are too small to efficiently deflect the ion flow, and need to merge in order to reach required size \citep{Milosavljevic2006,Spitkovsky2008,Chang2008,Davis2013}. Once the appropriate number of merging events has been achieved, the filaments reach the size of the ion Larmor radius. Only then is the ion flow deflected enough for the shock to start forming. Assuming this happens at a time $\tau$ and following the reasoning explained in \cite{BS14}, the shock formation time will be $2\tau$ in 2D, and $3\tau$ in 3D.

\subsection{Saturation phase}
The electron Weibel instability grows first, amplifying the field from its fluctuation value up to nearly equipartition with the electrons \citep{BS13}. The saturation time is here given by
\begin{equation}\label{eq:pair_time}
\tau_{s,e}=\delta_e^{-1}\Pi \omega_{pe}^{-1},
\end{equation}
where $\Pi$ is the number of e-foldings of the instability and $\delta_e\omega_{pe}^{-1}$ its growth rate. By this time, the field has grown to nearly equipartition with \citep{Medvedev1999,SilvaApJ}
\begin{equation}
B_{s,e}^2=8\pi\gamma n_0 m_e c^2.
\end{equation}
From the cyclotron frequency $\omega_{B_{s,e}}=qB_{s,e}/\gamma m_e c$, we derive the size of the filaments at saturation, which is also the electronic Larmor radius in $\mathbf{B}_{s,e}$ (we set $v_0 \sim c$),
\begin{equation}\label{eq:Le}
L_{s,e}= \frac{c}{\omega_{B_{s,e}}} = \sqrt{\frac{\gamma}{2}}\frac{c}{\omega_{pe}}.
\end{equation}
A full quantitative understanding of the saturation magnetic field is still lacking and will be explored elsewhere \citep[][]{SL15}.

Contrary to the electron current filamentation instability, which has to amplify the plasma thermal fluctuations, the ion Weibel instability finds an unstable mode already seeded, and further amplifies it, growing preferentially at that wavenumber. This is clear from the fact that the growth rate of the ion Weibel instability depends slowly on \(k\) for wavenumbers \(k \geq \frac{\omega_{pe}}{c} \left[ 1+ (\frac{m_e}{4 m_i})^{1/3} \right]^{-3/2}\) \citep[][]{D72}. At saturation, the ion Weibel instability has grown the field to
\begin{equation}
B_{s,i}^2=8\pi\gamma n_0 m_i c^2,
\end{equation}
while the size of the filaments is still given by Eq. (\ref{eq:Le}). The growth time of the ion phase of the instability is
\begin{equation}\label{eq:tausi}
\tau_{s,i}=\delta_i^{-1}\ln\left(\frac{B_{s,i}}{B_{s,e}}\right) \omega_{pi}^{-1} = \delta_i^{-1}\ln\sqrt{\frac{m_i}{m_e}} \omega_{pi}^{-1}.
\end{equation}
The total duration of this saturation phase is $\tau_s = \tau_{s,e}+\tau_{s,i}$, that with $\delta \equiv \delta_e = \delta_i \sqrt{m_i/m_e}$ we obtain
\begin{equation}
\tau_s = \delta^{-1}  \ln\sqrt{\frac{m_i}{m_e}} \left( \frac{\Pi \sqrt{\frac{m_e}{m_i}}}{ \ln\sqrt{\frac{m_i}{m_e}}} + 1 \right)\omega_{pi}^{-1}.
\end{equation}
Considering a number of e-foldings $\Pi \sim 20$, we find that even for an artificially low mass ratio of 100 the term \(\Pi \sqrt{m_e/m_i} / \ln\sqrt{m_i / m_e} \ll 1\). As a consequence, the duration of the whole saturation phase simply reads $\tau_s \sim \tau_{s,i}$. In reality, the ion growth rate \(\delta_i\) is lower than \(\delta_e \sqrt{m_e/m_i}\), the growth rate of the cold ion current filamentation instability, because it grows over a background of hot electrons. Yet, the consequences are negligible, see section \ref{sec:sftime}.

\subsection{Filaments merging phase}
The dynamics of the merging of the filaments has been studied previously in \cite{MedvedevApJ2005} who considered a simple 2D model of infinitely long cylindrical filaments of radius $D/2$, spaced by the distance $D$\footnote{The notation in this paper differs from the one used in \cite{MedvedevApJ2005}. The correspondence is obtained replacing the quantity $I_0/\sqrt{\mu_0}$, found in \cite{MedvedevApJ2005}, by $D\omega_{pi}v_0/4\sqrt{\gamma}$.}.

The filaments need to merge $n$ times in order to grow from the electronic Larmor radius (\ref{eq:Le}) to the ionic one. According to \cite{MedvedevApJ2005}, the expression for the merging time is different whether the transverse motion is relativistic or not (merging of the filaments implies a motion transverse to the flow). At any rate, the maximum transverse velocity achieved during one merging reads \citep[][]{MedvedevApJ2005}
\begin{equation}
v_m = \frac{1.67}{4}\frac{D \omega_{pi}}{\sqrt{\gamma}},
\end{equation}
considering $v_0\sim c$. We now compare this velocity to $c$. For the merging motion to be non-relativistic, we would need,
\begin{equation}
v_m \ll c  \Rightarrow D \ll  2.4 \sqrt{\gamma}\frac{c}{\omega_{pi}}.
\end{equation}
If this condition is fulfilled at the end of the merging phase, then it is always fulfilled, since filaments are growing with time. Replacing therefore $D$ by the ion Larmor radius $L_i=\sqrt{\gamma/2}c/\omega_{pi}$, we find the condition above is always satisfied.

For the regime of non-relativistic transverse merging, \cite{MedvedevApJ2005} established that all merging events are taking approximately the same time
\begin{equation}
\tau_0 = 2^{3/2} \sqrt{\gamma} \omega_{pi}^{-1}.
\end{equation}
This number now has to be multiplied by the total number $n$ of merging events. Since the initial and final filaments size are the electrons and ions Larmor radii respectively, $n$ is given by \citep{MedvedevApJ2005},
\begin{equation}
L_i = 2^{n/2} L_e \Rightarrow n = 2 \frac{\ln(L_i/L_e)}{\ln 2}= \frac{\ln (m_i/m_e)}{\ln 2}.
\end{equation}
The total duration of the merging phase is therefore given by
\begin{equation}\label{eq:merg}
\tau_m = n \tau_0  = \frac{2^{3/2}}{\ln 2} \ln (m_i/m_e) \gamma^{1/2}\omega_{pi}^{-1}.
\end{equation}

\subsection{Shock formation time}\label{sec:sftime}
We can now proceed to the calculation of the full shock formation time. As soon as the filaments are large enough, the ions are efficiently deflected by the field over a distance close to the ion Larmor radius. This happens at a time $\tau_s + \tau_m$. Assuming then $\delta^{-1}=\sqrt{\gamma/2}$ \citep{BretPoPReview}, we find,
\begin{eqnarray}
\tau_s + \tau_m &=& \left(\frac{1}{2^{3/2}}  + \frac{2^{3/2}}{\ln 2} \right)\gamma^{1/2}\ln \frac{m_i}{m_e}\omega_{pi}^{-1}\nonumber \\
                &\simeq& 4.43 \gamma^{1/2}\ln \frac{m_i}{m_e}\omega_{pi}^{-1}.
\end{eqnarray}
We thus find that the merging time is longer than the ion Weibel saturation time by a factor $2^3/\ln2 \sim 11.5$. As a consequence, the uncertainty on the ion Weibel growth rate does not affect significantly the formation time. In order to find out whether or not the flow is stopped in the overlapping region by $\tau_s + \tau_m$, we need to compare $L = 2c(\tau_s + \tau_m)$, the size of this region at this time, with the ion Larmor radius $L_i$. It is straightforward to show that
\begin{equation}
L/L_i = 8.83 \sqrt{2}  \ln \frac{m_i}{m_e}  \gg 1.
\end{equation}
As a consequence, the incoming flow is stopped in the overlapping region, and the shock starts forming. The downstream density at time $\tau_s + \tau_m$ is still only twice the upstream density. In order to reach the expected density jump of 3 for the 2D case\footnote{See \cite{SF12} and \cite{BretJPP2015} for the validity of the Rankine-Hugoniot jump conditions in collisionless plasmas.}, the system needs to evolve another $\tau_s + \tau_m$ since the overlapping region no longer expands \citep{BS14}. In the 3D case, the expected density jump is \(\approx 4\) so that waiting another $2(\tau_s + \tau_m)$ is required to bring enough material in the central region. Note that the present 2D model of the filaments merging as been successfully tested against 3D PIC simulation \citep{MedvedevApJ2005}. The shock formation time $\tau_{f,i}$ finally reads,
\begin{equation}\label{eq:ei_time}
\frac{\tau_{f,i}}{d} = 4.43 \gamma^{1/2}\ln \frac{m_i}{m_e}\omega_{pi}^{-1},
\end{equation}
where $d$ is the dimensionality of the system (\(d = 2 \, (3)\) for a 2D (3D) setup). Noteworthily, the expression does not reduce to the one obtained for a pair plasma by setting $m_e=m_i$  \citep{BS13,BS14}. The reason for this is that here we simply neglect the electron instability phase in the end result. Finally, we note that shocks in electron-ion plasmas form much slower than in pair plasmas for two reasons: on the one hand, the instability mechanism is slower while on the other hand, the merging phase is negligible in pair plasmas because the instability already generates filaments and fields large enough to deflect the flow.

\subsection{Comparison with the pair shock formation time}
For comparison of the shock formation times of pair and electron-ion shocks, Eq. (\ref{eq:pair_time}) gives the formation time $\tau_{f,e}$ for the pair case \citep{BS14}
\begin{equation}
\tau_{f,e}=d\tau_{s,e}=d\delta_e^{-1}\Pi \omega_{pe}^{-1}
\end{equation}
in terms of the dimension $d$ of the problem, whereas from Eq. (\ref{eq:ei_time}) we obtain for electron-ion shocks
\begin{equation}
\frac{\tau_{f,i}}{\tau_{f,e}} = \frac{6.2}{\Pi}\ln (m_i/m_e)\sqrt{m_i/m_e}.
\end{equation}
Considering $\Pi\sim 12$, which is the value obtained in the pair shock simulations presented in the next section, we find that an electron-ion shock forms 60 times slower than a pair shock for a mass ratio of 400, and 166 slower for a realistic mass ratio. For a mass ratio of 400, $\tau_{f,i}  = 3\sqrt{m_i/m_e} \tau_{f,e}$ whereas $\tau_{f,i} = 2.4\sqrt{m_i/m_e} \tau_{f,e}$ for a mass ratio of 100.

\section{Discussion of the results}\label{discussion}

We present now a series of 2D PIC simulations in order to validate the theoretical model. The counterstreaming beams were simulated in a simulation box with perfectly reflecting wall in the longitudinal direction and periodic boundary conditions transversally. The bulk is propagating along the \(x_1\) axis with Lorentz factors \(\gamma_0 = 25 - 10^3\), mass ratios \(m_i/m_e = 50-400\) and \(\mu = 10^6 \gamma_0\). The simulation box dimensions are \(L_x = 450  \, \sqrt{\gamma_0} c/\omega_{pe}\) and \(L_y = 150 \, \sqrt{\gamma_0} c/\omega_{pe}\) with \(\Delta_x = \Delta_y = 0.05 \, \sqrt{\gamma_0} c/\omega_{pe}\).

\begin{figure}[h!]
\begin{center}
\includegraphics[width=7cm]{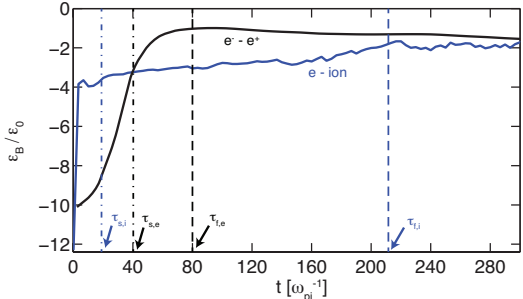}
\vspace{-24pt}
\end{center}
\caption{Normalised magnetic energy density for an electron-ion shock with \(m_i/m_e = 400\) (blue) and a pair shock (black) with \(\gamma_0 = 25\). The lines indicate the saturation times of the filamentation instability and the shock formation times.}\label{fig1}
\end{figure}

The magnetic field energy density is plotted in Fig.\ \ref{fig1} in order to get information about the role of the filamentation instability as mediator of the shock formation. This was done for a small slab along \(x_1\) in a region close to the wall, which will later be a region in the far downstream.

In the case of the pair shock the linear phase of the instability -- where the magnetic field grows exponentially in time and the magnetic field energy density is \( \epsilon_B \propto \exp(2 \delta_e t)\) -- can be clearly distinguished from the non-linear phase where the magnetic field has saturated. The growth rate of the cold electron instability fits very well the theoretical value \(\delta_e = 0.28 \, \omega_{pe}\), as well as the saturation field \(B_{f,e} \approx 7\, m_e c \omega_{pe}/ e\). The predicted saturation time of the filamentation instability of the pair shock \(\tau_{s,e} = 40 \, \omega_{pe}^{-1}\) and steady state shock formation time \(\tau_{f,e} = 80 \, \omega_{pe}^{-1}\) are shown in Fig.\ \ref{fig1} and match well with the simulation data.

In the case of the electron-ion shock with \(m_i/m_e = 400\), the evolution of the magnetic energy density shows several stages. The ion Weibel instability grows slower than \(\delta_i = \sqrt{m_e/m_i} \delta_e = 0.014 \, \omega_{pe} \) due to the influence of the hot electron background \citep{SS12}, which is negligible when compared to the full formation time. The final magnetic field at saturation of the ion instability matches approximately the theoretical value \(B_{f,i} \approx  140 c m_e \omega_{pe}/e\) as well as the saturation time in the simulation \(\tau_{s,i} = 18 \, \omega_{pi}^{-1}\). The theoretical model predicts a saturation time of \(13 \, \omega_{pi}^{-1}\), see Eq.\ (\ref{eq:tausi}).

\begin{figure}[h!]
\begin{center}
\includegraphics[width=7.5cm]{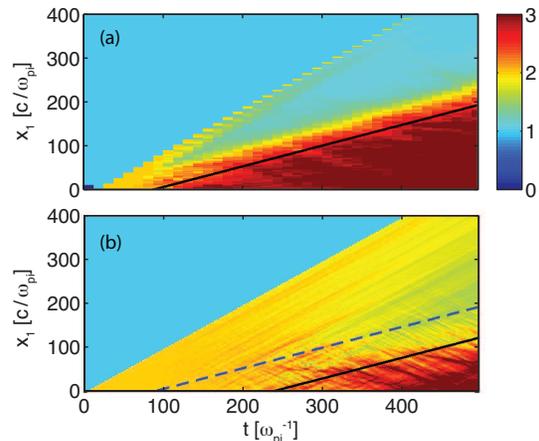}
\vspace{-24pt}
\end{center}
\caption{Ion density against space \(x_1\) and time \(t\) in units of \(\omega_{pi}^{-1} = \sqrt{m_i/m_e}\, \omega_{pe}^{-1}\) for \(\gamma_0 = 25\), \(m_i/m_e = 1\) (a) and \(m_i / m_e = 400\) (b). The black line indicates the shock front with \(n_2/n_1 = 3\) and the pair shock for comparison (dashed blue).}\label{fig2}
\end{figure}

\begin{figure}[h!]
\begin{center}
\includegraphics[width=7.5cm]{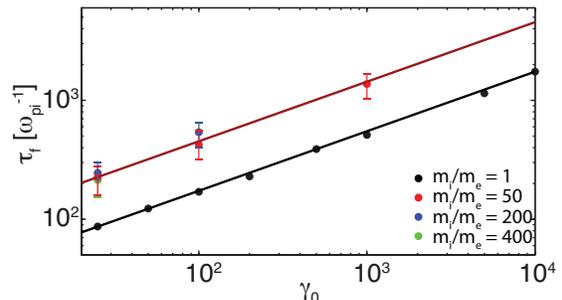}
\vspace{-24pt}
\end{center}
\caption{Shock formation time \(\tau_{f,i}\) vs. \(\gamma_0\) for a 2D pair shock with \(m_i/m_e = 1\) (black) \citep[][]{BS14} and electron-ion shock with different mass ratios (red). The error bars determine the uncertainty due to the finite size of the shock front.}\label{fig3}
\end{figure}

The black lines in Fig.\ \ref{fig2} indicate the shock front of the steady state shock with a jump of \(n_2/n_1 =3\). This line is extrapolated to \(x_1=0\) in order to define the shock formation time \(\tau_f\) as the intersection with the time axis. For the pair shock (Fig.\ \ref{fig2}a) it matches the theoretical value \(\tau_{f,e} = 2 \tau_{s,e}\), while for the electron-ion shock we observe \(\tau_{f,i} = 226 \, \omega_{pi}^{-1} \gg 2 \tau_{s,i} = 36 \, \omega_{pi}^{-1}\) (Fig.\ \ref{fig2}b). A systematic study with different velocity and mass ratio parameters provided a factor of \(2.5 \, \sqrt{m_i/m_e} \) with respect to the shock formation of pair shocks, which is consistent with Eq.\ (\ref{eq:ei_time}) (see Fig.\ \ref{fig3}). The shock width imposes an uncertainty in the determination of the shock formation time, which is represented by the error bars in Fig.\ \ref{fig3}.

\begin{figure}[h!]
\begin{center}
\includegraphics[width=3.7cm]{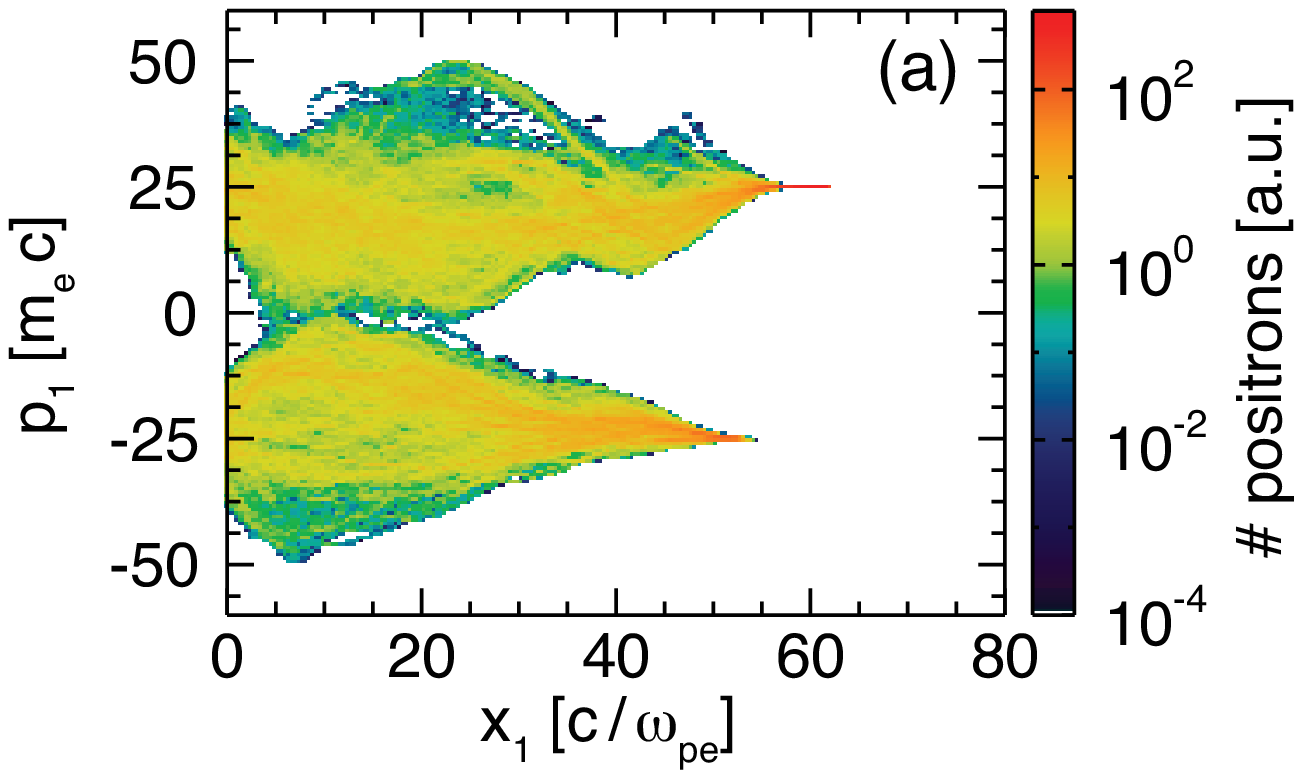}
\includegraphics[width=3.7cm]{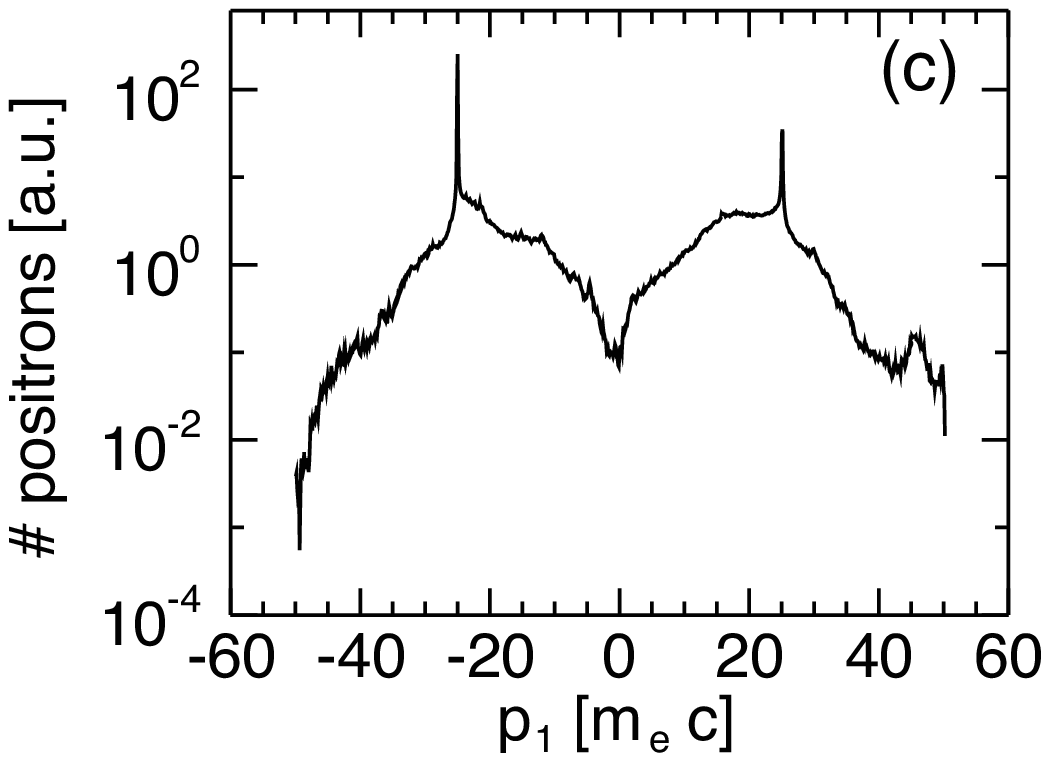}\\
\includegraphics[width=3.7cm]{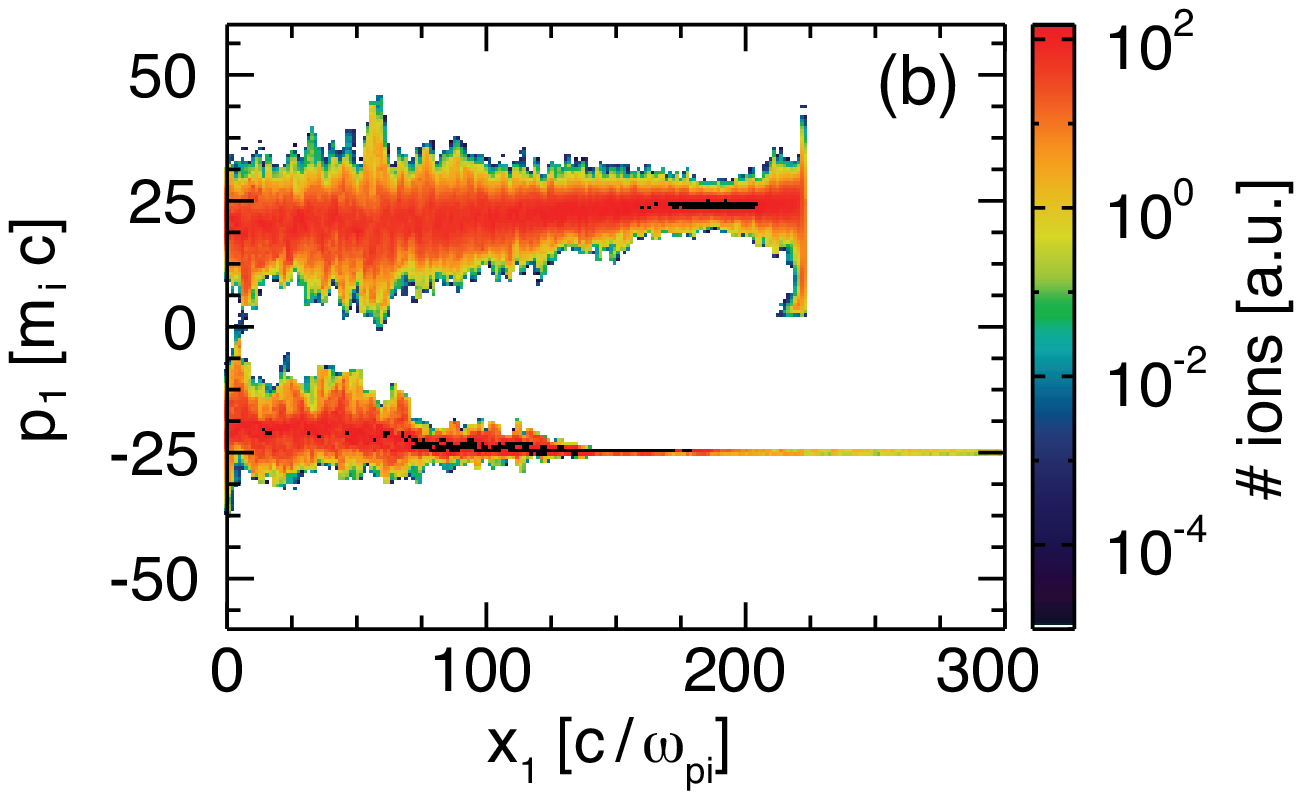}
\includegraphics[width=3.7cm]{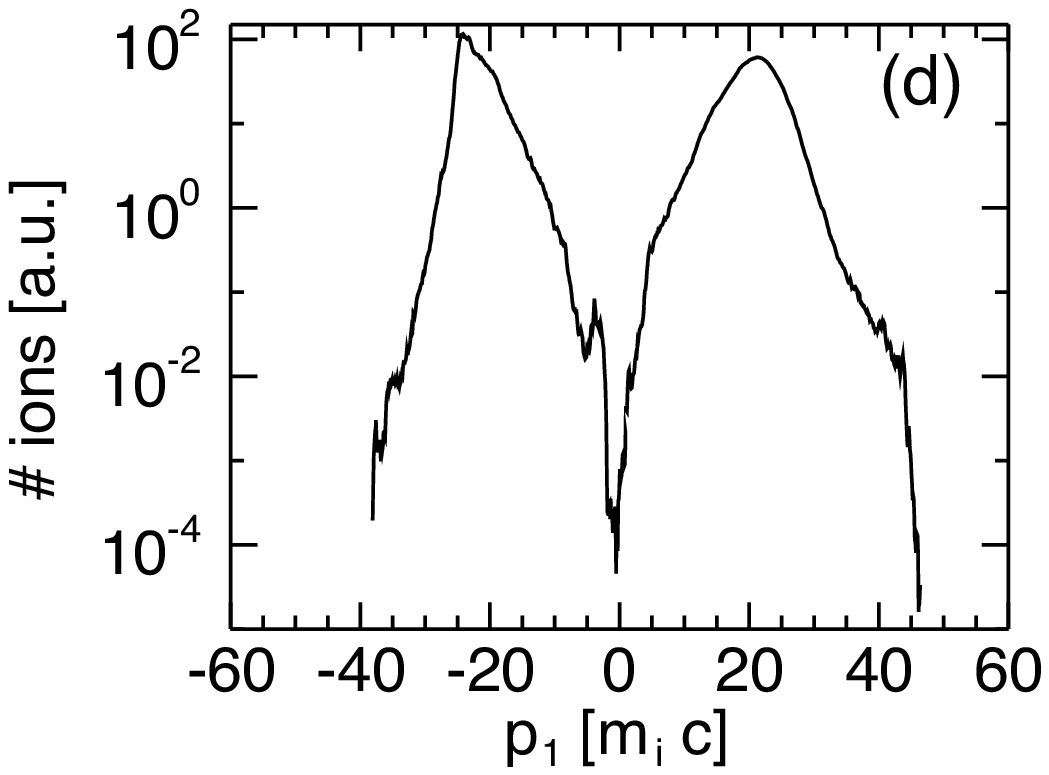}
\vspace{-24pt}
\end{center}
\caption{Momentum phase spaces of a pair shock at \(65 \, \omega_{pe}^{-1}\) (a) and electron-ion shock with \(m_i/m_e= 400\) at \(226 \, \omega_{pi}^{-1}\) (b) with respective averaged particle distributions (c) and (d).}\label{fig4}
\end{figure}

\begin{figure}[h!]
\begin{center}
\includegraphics[width=3.7cm]{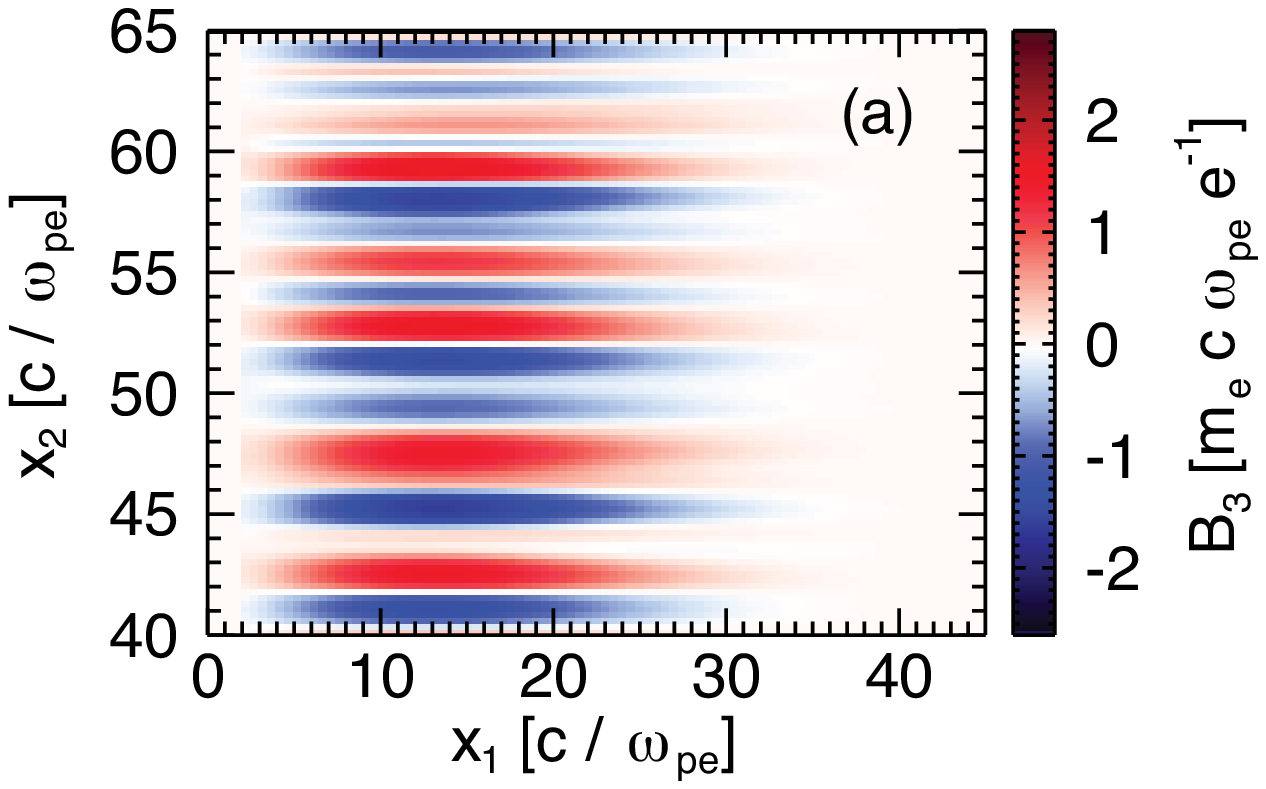}
\includegraphics[width=3.7cm]{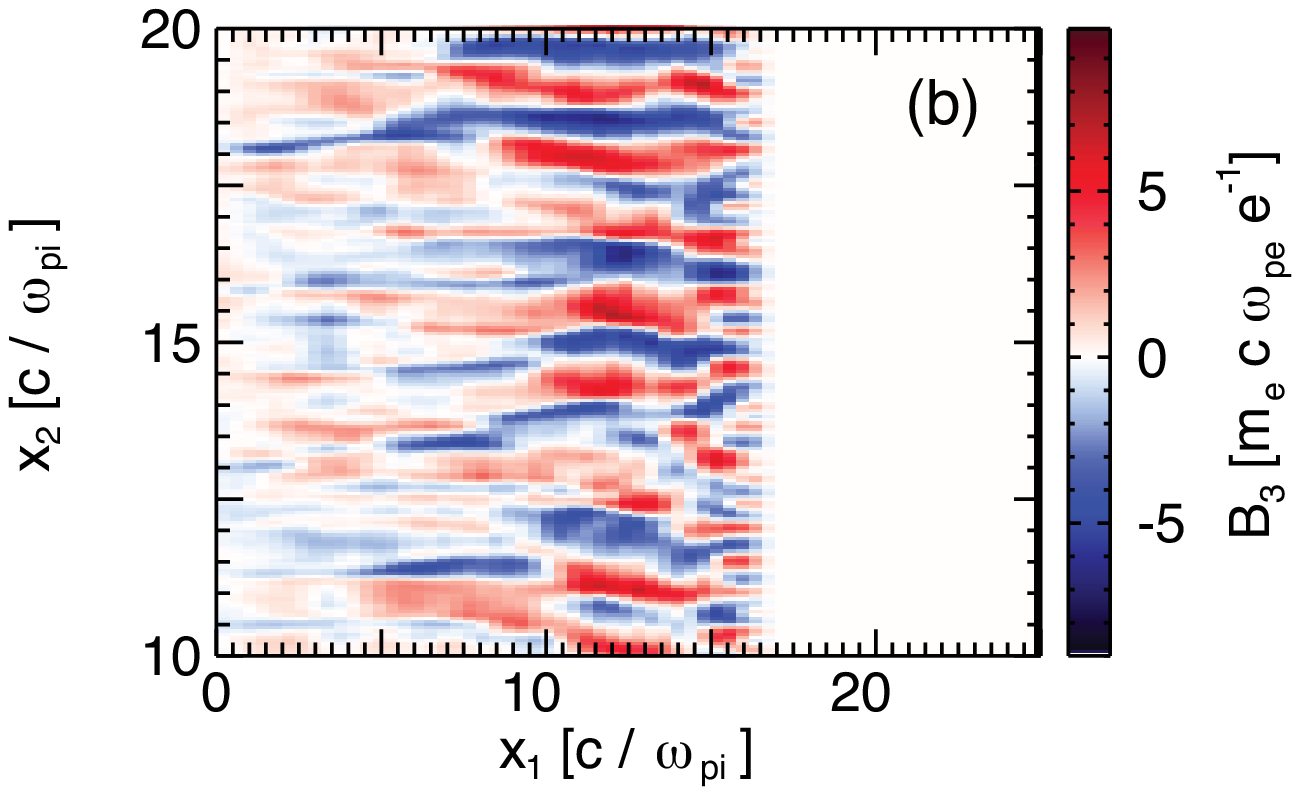}
\vspace{-24pt}
\end{center}
\caption{Magnetic field of the pair shock at saturation time \(\tau_{s,e} = 40 \, \omega_{pe}^{-1}\) (a) and for electron-ion shock at \(\tau_{s,i}=18 \, \omega_{pi}^{-1}\) (b).}\label{fig5}
\end{figure}

The delayed shock formation process in electron-ion shocks due to the merging of the filaments is demonstrated in Figs.\ \ref{fig4} and \ref{fig5}. In both cases, the accumulation of particles becomes very effective at the time when the phase space of the different beams starts to mix (Fig.\ \ref{fig4}). The first ions start to recirculate and change the sign of their longitudinal momentum. In panels (c) and (d) of Fig.\ \ref{fig4} this is demonstrated by a small fraction of particles close to zero momentum. At this stage the magnetic field turbulent scales are large enough that the ions can finish at least half a gyro circle, and the accumulation of particles becomes efficient. In pair shocks this process happens already right after the saturation of the filamentation instability, at \(t \omega_{pe} = 65\), whereas in electron-ion shocks this process takes much longer. For a mass ratio of \(m_i/m_e=400\), the efficient gyro reflection was observed only at \(t \omega_{pi} = 226\), which is slightly before the steady-state shock has formed.

An analysis of the magnetic field structure confirms this result. At the time when the magnetic field saturates, the transverse size of the magnetic field flux tubes is still too small in order to scatter the ions efficiently. The particles will feel the impact of another flux tube long before they can finish a gyro circle. For pair plasmas (Fig.\ \ref{fig5}a) the magnetic filaments at saturation time \(\tau_{s,e} = 40 \omega_{pe}^{-1}\) have reached a transverse spread of 4 \(c/\omega_{pe}\), while the maximum magnetic field strength is of the order of \(B_3 = 3 \,m_e c \omega_{pe}/e\). Particles with Lorentz factors \(\gamma \leq 12\) have Larmor radii that fit into this scale, meaning that they can recirculate in the field before being deflected by a different flux tube. This situation is different for electron-ion plasmas. For comparison, we plotted the magnetic field structure at \(\tau_{s,i} = 18\, \omega_{pi}^{-1}\) in Fig.\ \ref{fig5}b. At this stage, the magnetic field filaments show a transverse size on the order of 2 \(c/\omega_{pi}\) while \(B_3 \approx 5\, m_e c \omega_{pe}/e\), so that only particles with non-relativistic Lorentz factors \(\gamma \sim 1 \) will have Larmor radii on the same scale as the magnetic field tubes.  The filaments have to undergo a further merging process as described in Eq.\ (\ref{eq:merg}) until the transverse filament size becomes of the size of the ion Larmor radius for particles with \(\gamma > 25\).

\section{Conclusions}
We have investigated the full shock formation process in electron-ion plasmas in theory and simulations. In contrast to electron-positron pair shocks, where the shock formation time was found to be twice the saturation time of the filamentation instability, \(\tau_{f,e} = 2 \tau_{s,e}\), for electron-ion shocks the process is delayed to approximately \(\tau_{f,i} = 3 \sqrt{m_i/m_e} \tau_{f,e}\). The shock formation time is a sum of the saturation time of the instability plus an additional merging time, coming from the merging of filaments to the ion Larmor radius. At the time of the saturation of the instability, the filament size is still on the order of the electron Larmor radius. An extra time \(\tau_m\) is thus necessary for the condition for shock formation to be met, which is not the case for pair shocks. We applied the theory by \citet[][]{MedvedevApJ2005} to predict the merging time \(\tau_m\). The merging time retrieved from 2D PIC simulations is in agreement with theory and the shock formation time was confirmed to be \(2 (\tau_{s,i} + \tau_m)\). Slightly before this time, the first recirculation of ions was observed. At this stage, the scale of magnetic turbulence is large enough to trigger the density compression that precedes the full shock formation.

The different scales of the magnetic turbulence in electron-ion shocks compared with pair shocks, might have consequences for the particle acceleration process. This will be investigated in our following project.


\section*{Acknowledgements}
This work was supported by the European Research Council (ERC-2010-AdG Grant 267841), grant ENE2013-45661-C2-1-P from the Ministerio de Educaci\'on y Ciencia, Spain and grant PEII-2014-008-P from the Junta de Comunidades de Castilla-La Mancha. The authors acknowledge the Gauss Centre for Supercomputing (GCS) for providing computing time through the John von Neumann Institute for Computing (NIC) on the GCS share of the supercomputer JUQUEEN at J\"ulich Supercomputing Centre (JSC).

\end{document}